\documentclass[a4,10pt]{article}
\usepackage{graphicx}
\usepackage{multicol}
\usepackage{setspace} 
\usepackage[margin=2cm]{geometry}

\begin{document}

\begin{center}
\huge
\textsc{A brief guide to FXCOR}

\small

\vspace{2mm}

Mehmet Alpaslan

\emph{SUPA, School of Physics and Astronomy, University of St Andrews, St Andrews, KY16 9SS, U.K.}

\vspace{2mm}

July-September 2009

\emph{Departamento de Astronom\'{i}a y Astrofis\'{i}ca Facultad de F\'{i}sica, P. Universidad Cat\'{o}lica de Chile, Casilla 306, Santiago 22, Chile}

\end{center}

\onehalfspacing
\normalsize
\vspace{2mm}
\large
\noindent \textbf{1. Introduction and basic use}
\normalsize
\vspace{2mm}

FXCOR (which stands for Fourier cross-correlation) is a task within the software package IRAF, developed and distributed by the National Optical Astronomy Observatories (or NOAO) in Tucson, Arizona. It is mantained and supported by the IRAF programming group, and the software can be obtained free of charge at http://iraf.noao.edu/. There is plenty of documentation available for instruction on the basic use of IRAF and the various commands that are necessary to make the most of the software package. The author assumes basic knowledge of IRAF use for the rest of this document. This guide has been written using IRAF version 2.14 running on cygwin. Any code written using \texttt{this font} should be entered verbatim into the IRAF command prompt.

FXCOR uses the Fourier cross-correlation method developed by Tonry \& Davis in the 1979 paper\footnote[1]{ Tonry, J. \& Davis, M. \emph{A survey of galaxy redshifts. I - Data reduction techniques.} Astronomical Journal \textbf{84}, 1511-1525 (1979).} (hereafter referred to as TD). Briefly, this method involves correlating one spectrum of unknown redshift and velocity dispersion (known as the `object' spectrum) with another, the `template' spectrum, of zero redshift and a known velocity dispersion (in most cases, the template spectrum is that of a star, so there is no velocity dispersion to speak of). The task returns values for the redshift of the object spectrum (in the form of \emph{cz}) from the location of the cross-correlation peak as well as the FWHM of the peak, which is related to the velocity dispersion $\sigma$.

FXCOR itself is a task in the RV package of IRAF. Alongside FXCOR, RV contains the following tasks: CONTINPARS, RVCORRECT, RVREIDLINES, FILTPARS, KEYWPARS and RVIDLINES. Of these, CONTINPARS, FILTPARS and KEYWPARS are of central importance when using FXCOR and will be discussed later. The help file for FXCOR also contains valuable information detailing the various parameters occupied by the task, as well as some examples on how to use it. This can be viewed by typing \texttt{help fxcor}. FXCOR itself is accessed by entering \texttt{rv} into the IRAF command prompt once the program has been started up. This also displays the other tasks within the RV package, as listed above. The RV package, like any other, can be exited with the \texttt{bye} command; and IRAF itself can be exited using \texttt{logout}.

As with any other task within IRAF, FXCOR is run by typing its name into the command prompt (or a minimum number of characters to identify it uniquely, such as \texttt{fx}). This will prompt the user to specify the object and template spectra and a very basic cross-correlation takes place.\footnote[2]{A minor complication arises here. IRAF \emph{must} be run while the user is in the folder containing the login.cl file and uparms folder. These contain vital information for IRAF to run. Once the program has been started, the user may navigate to the folder containing the spectra they wish to analyse.} This, however, is far from the most powerful way of using FXCOR, as it does not allow the user to manipulate many parameters for the task to use. Once again, as with any other task in IRAF, a more detailed version of the task can be run by typing \texttt{epar fxcor}.
	
\begin{center}
	\includegraphics[width=0.50\textwidth]{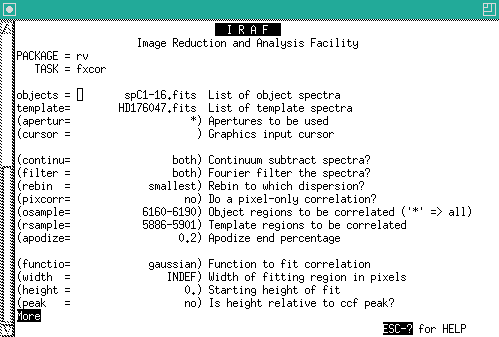}
\end{center}
	\footnotesize
	Figure 1. A screen capture of the parameters of FXCOR that can be changed after typing \texttt{epar fxcor}. These will be discussed with more detail later.
\vspace{4mm}
\normalsize

The user may navigate through these parameters by using the directional keys on their keyboard. Any changes to the parameters are written normally and verified by hitting Enter. Finally, when the user is satisfied, they must enter \texttt{:go} for the task to run. The help document for FXCOR that comes with IRAF explains, in detail, each of these parameters. Some of the more important ones will be discussed further on in this document. 

Some work must be done on the spectra to be cross-correlated before they can be worked on with FXCOR. To begin with, it is recommendable for both the object and template spectra to have the same resolution (wavelengths/pixel); but it is not absolutely necessary as FXCOR can be told to rebin one of the spectra to the resolution of the other (this is explained in the \texttt{rebin} parameter of FXCOR). It is also helpful for them to begin and end at the same wavelengths; however this requirement can be circumvented by selecting the zones to be cross-correlated carefully within FXCOR. Ideally, the range on the template should include notable absorption or emission features, and the range on the object spectra should cover those same features. Essentially, it is important for the task to not correlate zones with no data. The headers for some of the spectra must also be modified. To begin with, each spectra should have an observatory specified--this allows IRAF to properly calculate heliocentric corrections. This can be done using the \texttt{hedit} command with the \texttt{add=yes} option. The keyword that needs to be added is \texttt{OBSERVAT}. Similarly, the template spectra should be given the \texttt{VHELIO} header, with their heliocentric velocities given in units of km s$^{-1}$.

Once the parameters have been set and the program run with \texttt{:go}, the following screen is presented to the user:

\begin{center}s
	\includegraphics[width=0.50\textwidth]{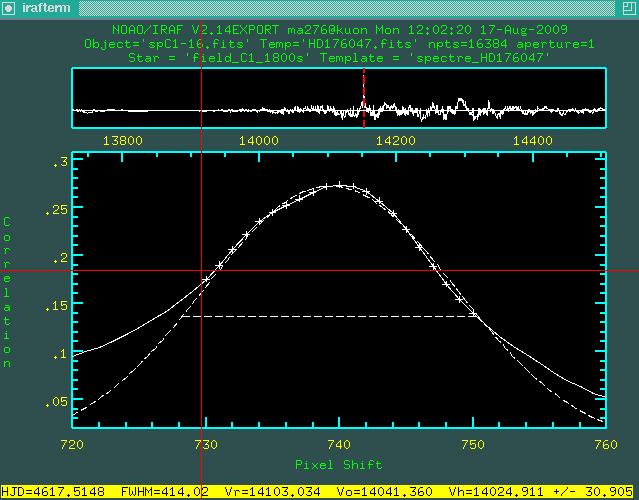}
\end{center}
	\footnotesize
	Figure 2. The main display of FXCOR after a cross-correlation has been completed. The plot in the upper section of the screen displays the entire cross-correlation function (CCF), with the red dashed lines indicating the vertical borders of the section that has been zoomed in on in the main plot below. The dashed lines are the fit (in this case a Gaussian) with its FWHM. The \emph{y}-axis represents the correlation height, and the \emph{x}-axis represents the pixel shift. Finally, in the yellow bar below, detailed information is given regarding the fit.
\vspace{4mm}
\normalsize

The goal for any good fit is for the cross-correlation to be as high as possible. The values on the \emph{y}-axis represent the correlation percentage. TD state that any value above 0.8 is `excellent' while a cross-correlation of about 0.5 is still considered good. Another important parameter for determining the quality of a fit is given by the Tonry \& Davis R-value, which is defined as the ratio between the height of the real peak and the rest of the CCF. Briefly, $R = h/\sqrt{2} \sigma_a$, where $h$ corresponds to the height of the true peak, and $\sqrt{2} \sigma_a$ corresponds to the average peak in the rest of the CCF. Ultimately, however, it is down to the user's common sense whether or not a value returned by the program should be accepted or rejected, but the correlation height and R value are good guidelines.

Once in this screen, the user has access to many different options. If they are unsatisfied with the fit, keystroke \texttt{g} allows them to specify the range, on the main plot, where points should be fitted to a line. This can be used, for example, to tell FXCOR to only fit points to the upper section of a CCF peak, instead of the entire thing. Keystroke \texttt{e} will move the user to an alternative view of the fit, zooming out a little bit and giving some more detailed information the fit (including the TD R value). Hitting \texttt{v} will enter the user into verbose mode, where a very detailed breakdown of the calculation will be presented on the IRAF terminal itself. This verbose printout contains information not only on the height of the fit, its R value and the FWHM, but also the wavelength range used, and a number of other pieces of useful information. Finally, keystroke \texttt{w} will save the results to a series of files (one \texttt{.txt} file containing a summary of results, one \texttt{.log} file containing the same content as the verbose mode results given by hitting \texttt{v} and finally, one \texttt{.gki} image file of the fit); and \texttt{q} quits the fit and returns the user to the IRAF command prompt.

Finally, the FWHM values are related to the velocity dispersion by the creation of a \emph{calibration curve.} This is detailed in Appendix A. Briefly, this is constructed by correlating the original template spectrum with itself, but convolved with a Gaussian of set width. By doing this several times, a calibration curve can be made, and a calibration function interpolated with it.

\vspace{2mm}
\large
\noindent \textbf{2. Parameters and KEYWPARS}
\normalsize
\vspace{2mm}

There are various parameters in FXCOR that can be modified with the \texttt{epar fxcor} command. Below is a brief description of the most important ones. Remember that the IRAF help file for FXCOR explains all of these in great detail. The program assigns default values to the parameters initially, and these should only be modified if very specific analysis needs to be done. In particular, one tends to focus on which files the task should look at, the type of fit that should be made, and the spectral range that the task must perform the correlation.

\begin{itemize}
	\item \texttt{object} and \texttt{template}: Specify the object and template files you wish to correlate here, and separate them with commas.
	\item \texttt{apertur}: List of apertures to be correlated in echelle spectra. If left at the default value of *, it will correlate all the apertures.
	\item \texttt{continu}: This tells FXCOR to apply continuum substraction to either the \textbf{o}bject, \textbf{t}emplate, \textbf{n}one or \textbf{b}oth. This, along with filtering, will be explained in the next section.
	\item \texttt{filter}: As with continuum, this tells FXCOR which spectra to apply a filter to.
	\item \texttt{rebin}: This tells FXCOR how to rebin the data, if this is necessary. This is only important if the input spectra have different dispersions. Counter-intuitively, it may sometimes be better to rebin to the highest resolution (worst dispersion) as in this case FXCOR will be downgrading the quality of one spectrum as opposed to trying to artifically improve another.
	\item \texttt{pixcorr}: This tells FXCOR whether or not to do a pixel-only correlation. This will neglect velocity information in the output, and the FWHM will only be given as a function of pixels. This is useful when the user is interested in the pixel shift information.
	\item \texttt{osample} and \texttt{rsample}: Object and template region to be correlated. The input is given in the format of $N-n$ where $N$ and $n$ are two different numbers corresponding to the starting wavelength and the ending wavelength. If you wish to specify these ranges in pixels, then put p in front of the number you wish to use. For example, \texttt{osample 4400-5050} will correlate between 4400 and 5050\AA but \texttt{osample p4400-p5050} will correlate between pixel number 4400 and 5050. Selecting these carefully is very important.
	\item \texttt{function}: This chooses the type of function that is fit to the data. Details on the types of function can be found in the help file.
	\item \texttt{observa}: If the observatory is not specified in the FITS header of the spectra, then it can be specified here instead. The observatory name can be added from the observatory task in IRAF. It is usually a small code, and can be found online. One good database is http://tdc-www.harvard.edu/iraf/rvsao/bcvcorr/obsdb.html. The help file for the observatory task gives information as to how to add new observatories to the database on your computer.
\end{itemize}

Very briefly, KEYWPARS serves a specific purpose. FXCOR extracts a lot of data from the header of the FITS files used for correlation. It recognises which data to retrieve by looking at the labels of the data lines of the header, such as OBSERVAT and VHELIO. However, sometimes there are differences in naming conventions between the FITS header and what FXCOR is looking for. So, KEYWPARS is used to tell FXCOR what everything is labelled as in the FITS files it is told to correlate, such as RA and Dec, observed velocity, HJD and etc. All of these can be changed by executing the KEYWPARS task in the RV package. This is a fairly self-contained task, and when it is run, it will pull up a list of parameters similar to Fig. 1 which can be navigated using the arrow keys. To save, simply hit CTRL-D and to exit without saving, CTRL-C. This input applies to all IRAF tasks.

\vspace{2mm}
\large
\noindent \textbf{3. Continuum subtraction and filtering}
\normalsize
\vspace{2mm}

It is important to subtract the continuum from both spectra before the cross-correlation takes place, in order to reduce the potential correlating the wrong peak (for example, a coincidental peak in the CCF created by noise). Furthermore, filtering both spectra gets rid of high frequency noise, leaving only the most useful data. There is no mention in existing literature that filtering negatively affects results by widening correlation peaks. Both of these procedures are done according to the parameters set in the CONTINPARS and FILTPARS tasks in the RV package. 

To edit the continuum subtraction parameters, the user must called up the \texttt{continpars} task. As with modifying parameters for FXCOR, the user can navigate through the various parameters and change them at their leisure. However, instead of typing \texttt{:go} into the command line, they must save and quit by hitting CTRL-D. Hitting CTRL-C exits the task without saving.

Continuum subtraction is an operation that IRAF can do on its own, and quite effectively at that. It uses the ICFIT task in order to do this, and the help file for ICFIT describes its parameters in great detail. Simply specifying a continuum fitting function of order around 5 is usually enough. Order 2 is an absolute minimum.

Applying a filter to the spectra can be considerably more complicated, as there are many more options that need to be considered. FILTPARS is used to enter the parameters for filtering. Here, the user can choose which kinds of filters to apply to the spectra, and what through what ranges (specified in this case in terms of wave number). Detailed information on the kinds of filters that can be applied is found in the FILTPARS help file. Most of the time, the ramp filter seems to be the most useful. 

The filter is described by four wavenumbers: the cuton value $k_1$ where the filter begins to rise from 0, until it reaches 1 at the fullon value $k_2$. It continues at this level until the cutoff value, $k_3$, and back to 0 at $k_4$. To get rid of low-frequency residual continuum information, a value of $k_1 = 4-8$ and $k_2 = 9-12$ should be applied. $k_3 = N_{pix}/3$ and $k_4 = N_{pix}/2$ are good values for the last two wavenumbers. $N_{pix}$ refers to the number of pixels in the object spectra. Wegner et al. in their paper detailing the dynamics of early-type galaxies\footnote[1]{Wegner, G. et al. \emph{The peculiar motions of early-type galaxies in two distant regions - II. The spectroscopic data.} Monthly Notices of the Royal Astronomical Society \textbf{305}, 259-296 (1999).} give a similar set of values to use. 

\vspace{2mm}
\large
\noindent \textbf{4. General tips}
\normalsize
\vspace{2mm}

\begin{itemize}
	\item It is advisable at first to do a cross-correlation using the entire spectral range in order to determine the redshift of the object. Then, by specifying ranges using the parameters described below, around known absorption and emission lines, specific cross-correlation peaks can be found, and from them, FWHM values can be obtained.
	\item Often, the objects to be correlated will be of unknown redshift. The redshift can be determined through FXCOR, and this is then used to work out where some of the more notable absorption or emission likes are located. The command \texttt{:printz y} converts the velocity measurements in the status bar of FXCOR to redshifts. Hitting \texttt{x} will redo the fit and yield redshifts instead of velocities.
	\item When it comes to elliptical galaxies, most of the literature tends to correlate around the \emph{Mg} triplet, the \emph{Na$_D$} line, and \emph{H}$\beta$ line.
	\item A helpful way of identifying absorption and emission lines is by using the program SPLAT-VO from the Starlink package, found at http://star-www.dur.ac.uk/$\sim$pdraper/splat/splat-vo/. This software allows users to include line identifiers over spectra, and shift these identifiers according to redshift. So, by determining the redshift of the object using FXCOR and correlating the entire spectrum range, one can identify absorption and emission lines in the object spectrum using SPLAT. Then, these ranges can be specified in the FXCOR parameters and used for specific velocity dispersion cross-correlation measurements.
	\item Intuitively, if the user switches the object and template spectra around, the results are similar. The FWHM stays the same, but the redshift becomes negative.
	\item It is possible to give FXCOR a list of object and template spectra to cross-correlate. This is particularly useful if you need to correlate a lot of data, and are comfortable enough with your data to do it automatically.
\end{itemize}

\vspace{2mm}
\large
\noindent \textbf{5. Acknowledgements}
\normalsize
\vspace{2mm}

I would like to thank the Student-Staff Council of the School of Physics and Astronomy at the University of St. Andrews for funding the research grant that made this exchange possible, and naturally, to Professor Andreas Reisenegger at P. Universidad Cat\'{o}lica for taking me on as a summer research student, as well as extensive help and consultation during the project. I would also like to thank Hern\'{a}n Quintana and Paula Zelaya for all of their help.

\newpage

\large
\textbf{Appendix 1: Calibration curves relating FWHM and $\sigma$}
\normalsize
\vspace{2mm}

While one might be inclined, initially, to simply convert the FWHM of the correlation peak to $\sigma$ using a mathematical formula relating the FWHM of a Gaussian to its spread ($\sigma$ in $f(x)=\frac{1}{\sigma \sqrt{2 \pi}} \exp(\frac{-x-\mu^2}{2 \sigma^2})$, not to be confused with the velocity dispersion $\sigma$), this isn't necessarily the case when using FXCOR. The general consensus is that one must create a calibration curve using the template spectra at their disposal. This is done by convolving the template spectra with Gaussians of fixed with, followed by correlating these with the original template spectra. By plotting the FWHM of these correlations with the spread of the Gaussian, one can construct such a curve. An example is shown below:

\begin{center}
	\includegraphics[width=0.50\textwidth]{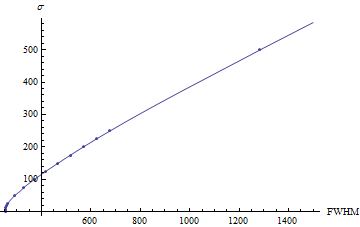}
\end{center}
	\footnotesize
	Figure 3. An example calibration curve for a template star correlated with galaxies. By interpolating the points on the plot, a function (the straight line in this plot) can be used to relate the FWHM of the correlation to the velocity dispersion. Both axes are in units of km s$^{-1}$.
\vspace{2mm}

\normalsize
It is important to note that these calibration curves should be made on a range identical to that of the correlation being done with the objects and templates. For example, if the correlation is being done on a range of 4700-5700\AA, then the calibration curve should be created using the same range. On the other hand, if the correlation is being done with individual lines, it will be necessary to create a separate calibration curve for each line used. 

The IRAF task used to convolve the original template spectra with a Gaussian is simply called GAUSS and can be accessed from the opening IRAF screen. The program will prompt the user for an input file and a name of the output file -- make sure that you do not overwrite your original spectra! Another important thing to note is that the GAUSS task will ask for the width of the Gaussian in pixels. This width should then be converted to kms$^{-1}$ by looking at the pixel-to-velocity resolution given in FXCOR's output (specifically, in the .log or .txt files).

I recommend creating a large number of convolved spectra, taking values for $\sigma$ that are mostly below 10 pixels. This is because at small values the function is harder to interpolate. I personally go from 0 to 1 in increments of 0.25, then from 1 to 10 in increments of 1, and then progress to the upper limit in increments of 10. This is because the calibration function can diverge at small values of $\sigma$, so it is best to take many points near that limit, to have it well-defined.

Once the points have been obtained, a basic mathematics software package like Mathematica or Maple can be used to interpolate the points and create a correlation function, into which the FWHM values can be fed, and velocity dispersion values can be obtained.

\end{document}